# AN AFFECTIVE AWARE PSEUDO ASSOCIATION METHOD TO CONNECT DISJOINT USERS ACROSS MULTIPLE DATASETS– AN ENHANCED VALIDATION METHOD FOR TEXT-BASED EMOTION AWARE RECOMMENDER

John Kalung Leung[1], Igor Griva[2] and William G. Kennedy[3]

[1]Computational and Data Sciences Department, Computational Sciences and Informatics, College of Science, George Mason University, 4400 University Drive, Fairfax, Virginia 22030, USA
`jleung2@gmu.edu`

[2]Department of Mathematical Sciences, MS3F2, Exploratory Hall 4114, George Mason University,4400 University Drive, Fairfax, Virginia 22030, USA
`igriva@gmu.edu`

[3]Center for Social Complexity, Computational and Data Sciences Department, College of Science, George MasonUniversity, 4400 University Drive, Fairfax, Virginia 22030, USA
`wkennedy@gmu.edu`

## ABSTRACT

*We derive a method to enhance the evaluation for a text-based Emotion Aware Recommender that we have developed. However, we did not implement a suitable way to assess the top-N recommendations subjectively. In this study, we introduce an emotion-aware Pseudo Association Method to interconnect disjointed users across different datasets so data files can be combined to form a more extensive data file. Users with the same user IDs found in separate data files in the same dataset are often the same users. However, users with the same user ID may not be the same user across different datasets. We advocate an emotion aware Pseudo Association Method to associate users across different datasets. The approach interconnects users with different user IDs across different datasets through the most similar users' emotion vectors (UVECs). We found the method improved the evaluation process of assessing the top-N recommendations objectively.*

## 1. BACKGROUND

We developed a contextual text-based Emotion Aware Recommender (EAR) in [1] as one of the five Recommended comparative platforms. We worked with movie metadata datasets; thus, each Recommender's primary task is to generate a top-N movie recommendations list that meets an active user's tastes and preferences. We advocated in [1] to develop an Emotion Aware Recommender by leveraging on the emotional profiles represented by movie emotion vectors (mvec) and user emotion vectors (uvec) from a contextual affective concept that we have initially developed in [2]. As we came to evaluate the top-N recommendations lists that generated by the five Recommenders in [1], we could only subjectively contrast the differences among the top-N recommendations lists based on our personal experience and sentiment. We acknowledged the deficiency of lagging an objective evaluation metrics to assess Recommenders' top-N recommendations lists accurately. We place a future work order in [1] to develop a proper evaluation method for assessing Recommender's top-N recommendations list.

Moreover, we observed that the ml-latest-small (hereafter a.k.a. ml-sm) MovieLens dataset used in [1] for training Recommenders in the comparative platform contains 100,836 ratings on 9,742 movies by 610 users. On average, each user rated about 170 movies. The actual rating number is much small in about the low 50s. For example, we picked ml-sm user id 400 as the test user who only rated 43 movies. Many users in the ml-sm dataset rated 20 plus to 50 plus movies. The frequency of rating movies is good enough for formulating the user emotion profile, uvec, for training but leaves fewer to none for testing and validation. In [1], we advocated the notation of combining ml-20m, ml-25m, and ml-latest-full (hereafter a.k.a. ml-27m) datasets for testing and validation. However, MovieLens does not maintain users' ids consistently across different datasets, but only within a dataset. We faced the problem of associating test user id 400 in the training dataset, ml-sm, to other three more massive datasets aiming for testing and validation. We want to determine which user or users in other MoiveLens datasets can link with the training user id 400 in the ml-sm dataset. By experimenting in [1], we are sure that user id 400 is not the same user found in different MovieLens datasets. We shall address how to associate users across different MovieLens datasets in this paper.

We make use of MovieLens datasets depict in Table 1 to develop the five Recommenders in the comparative platform, as illustrated in [1]. Each of the four MoiveLens datasets used for our study contains a rating data file composed of users' rating history of watched movies. Users in MovieLens are randomly selected. However, user ids are consistent between ratings and tags datafiles within a dataset, but not across datasets. For example, user ids in the ml-sm dataset are consistent and refer to the same users in the ratings and the tags data files within the dataset. However, user ids are not consistent across MovieLens datasets, such as across ml-sm and ml-20m datasets.

In our movie Recommender development, we choose the ml-sm dataset as the training dataset while combining the other three MovieLens datasets: ml-20m, ml-25m, and ml-27m, as testing and validation datasets. The ml-latest-small dataset contains 610 users, while the other three MovieLens datasets contain numbers of users in the range of 138,000 to 280,000, as depicted in Table 1. We want to find a way to associate users correctly among these datasets for the training, testing, and validation. By leveraging user emotion profile, uvec, we can associate users in training, testing, and validation datasets by connecting the most similar uvec among pairs of users. We named the scheme Affective Aware Pseudo Association (AAPA) method. To our best knowledge, we may be the first investigator party to apply such a method in affective computing to align disjoint users across different datasets.

## 2. INTRODUCTION

We have documented in [2] the affective embedding technique of applying the movie emotion profile and user emotion profile through movie emotion vectors (mvec) and user emotion vectors (uvec) as the affective computing embeddings for a Recommender to generate a top-N recommendation list during the recommendation making process. [1] aims to employ mvec and uvec embeddings as emotional components to develop an end-to-end Emotion Aware Recommender (EAR). As illustrated in [2], mvec represents a movie's emotional profile obtained through the emotion classification of movie overviews using a Tweets Affective Classifier (TAC) that we have developed in [2]. TAC can classify six basic human emotions, and we include a neutral emotion for affective computing convenience. For each scraped film, we apply TAC against movie overviews scraped from The Movie Database site to classify the movie emotion profile, mvec. To obtain a user's emotion profile, uvec, we compute the average of all watched films' mvecs by the user. In [1], we include other movies' textual metadata, such as genres in the mvec embeddings. We denote the expanded mvec as item vectors (ivec). We also named the extended uvec, wvec.

We demonstrated, in [2], the affective movie recommendation making through an SVD-CF Recommender. In [1], we build a comparative Recommender platform, making movie recommendations through five Recommender filtering algorithms. The filtering algorithms include Collaborative Filtering (CF) and Content-based filtering (CB). We developed a movie Genres Aware Recommender (GAR). We leverage TAC to classify the moods of mvec against the movie overviews and transform the movie mvec embeddings into a multi-label emotion classification in One-Hot Encoded (OHE) embeddings. We denoted the mvec expanded OHE embeddings as ivec. We applied the ivec embeddings in the development of the Emotion Aware Recommender (EAR). We combine the emotional aspect of EAR and genres aware of GAR into an expanded ivec embedding to develop a Multi-channel Aware Recommender (MAR). We also developed from scratch a User-based Collaborative Filtering (UBCF) Recommenders and an Item-based Collaborative Filtering (IBCF) Recommenders for the comparative platform. Through the Recommenders recommendations-making process in the comparative platform, we observed the performance and compared the differences among the five Recommenders.

We apply the Cosine Similarity depicted in equation 2 as the primary algorithm in building our Recommender platform. In the case of making top-N movie recommendations of similar genres that the active user has watched and liked, Cosine Similarity will reveal the closeness in the similarity between the recommended movies and the movies the active user has viewed and liked. Similarly, we can apply Cosine Similarity to find the similarity in the emotion profile of a top-N movies list and the movie's emotion profile of an active user who has watched and loved. Moreover, in UBCF, we apply a rating matrix, $R$, to compute the collaborative filtering for recommending movies to an active user. Each row of UBCF in $R$ represents the rating value of films a user has watched and rated; whereas, each column in $R$ represents a movie of rating scores it received from users who have viewed and assessed. By comparing the Cosine Similarity between the active user and a user in the corresponding rows, we effectively compare two rows in $R$; thus, we know the the two users' closeness. Once we find the closest similarity score of the active user and a particular user in $R$, we scan the active user's unwatched movies that match the closest similar user has watched films. Through collaborative filtering, we make the top-N movie recommendations to the active user. Lastly, we evaluate the performance of Recommenders in the comparative platform by contrasting each top-N recommendation list generated by the five Recommender algorithms. We find the top-N recommendation list made by the Emotion Aware Recommender (EAR) shows intrigue results.

In the advent of the Internet era, large conglomerates, small and medium businesses (SMB) have deployed Recommender Systems to gain a competitive advantage in providing customers a personalized, useful business transaction experience while understanding customers' tastes and decision-making habits. For customers who left feedback regarding their experiences of the goods and services they received, Recommender can mine customers' opinions through sentiment analysis (SA) to better understand the what, why, and how customers' likes and dislikes the goods and services they consumed. Also, suppose customers have rated the goods and services, Recommender can use the rating information and the sentiment analysis on opinion feedback to make a future personalized recommendation of products and services to customers that meet their tastes and expectations. For example, such Recommender is known as Hybrid Recommender System using Collaborative Filtering with Sentiment Analysis (CF-SA) [3]. CF-SA

Recommender is also known to outperform the baseline Collaborate Filtering Recommender System in personalized recommendation making [4] [5].

Nevertheless, no Recommender was built with design to explicitly collect human emotions data [6] [7]. Also, no publicly available dataset contains explicit affective features for implementing a Recommender System. The alternative for Recommender researchers is to build an affective aware Recommender by deriving the needed emotional features from some datasets implicitly [8] [9] [10]. Movies and music datasets are the two most popular datasets with metadata, such as genres and reviews for affective features mining [6].

We will describe the related work in the next section to level set the readers the works of affective computing and Emotion Aware Recommenders (EAR). In the methodology section, we will illustrate the development work of the five Recommender algorithms used in the comparative platform. We briefly describe the Tweets Affective Classifier (TAC) and datasets used. Next, we will highlight the five Recommenders in the implementation section. In the evaluation section, we will show the top-N recommendations lists generated by the five Recommender algorithms while contrasting their differences. We also will highlight our observations regarding the limitations and deficiencies of developing the comparative platform. We will document our future work plan in the future work section before closing our report with a conclusion. Following the conclusion is the reference section and the authors' brief biography.

## 3. RELATED WORK

Emotion Aware Recommender System (EAR) is still in active research. Illustrated below are samples of a few recent works. Orellana-Rodriguez [11] [12] advocated that instead of detecting the affective polarity features (i.e., positive/negative) of a given short video in YouTube, they detect the paired eight basic human emotions advocated by Plutchik [13] [14] into four opposing pairs of basic moods: joy–sadness, anger–fear, trust–disgust, and anticipation–surprise. Orellana-Rodriguez [11] also leveraged the auto extraction of film metadata's moods context for making emotion-aware movie recommendations. Qian et al. [15] proposed an EARS based on hybrid information fusion using user rating information as explicit data, user social network data as implicit information, and sentiment from user reviews as the source of emotional information. They [15] also claimed the proposed method achieved higher prediction ratings and significantly enhanced the recommendation accuracy. Also, Narducci et al. [16] [17] described a general architecture for building an EARS and demonstrated through a music Recommender with promising results.

Moreover, Mizgajski and Morzy [18] advocated an innovative approach for making recommendations on a large-scale news Recommender through a multi-dimensional model EARS. The database, based on 2.7 million unique user's self-assessed emotional reactions, resulted in over 160,000 emotional reactions collected against 85,000 news articles, consists of over 13 million news pages. Katarya and Verma [6] completed a literature review of research publications in the Affective Recommender Systems (ARS) field from 2003 to February 2016. The report offers in-depth views of the evolution of technology and the development of ARS.

Human primary Emotion Detection and Recognition (EDR) is still in active research [19] [20] [21] [22] [23] [24] [25] [26]. The main focus of Facial Detection and Recognition (FDR) is in image-oriented research [27] [28]. The main thrust is to detect basic human emotion through human facial expression. However, subjective writing takes the lead in textual based sentiment analysis (SA) [29] [30] [31]. The aim is to mine fine-grained sentiment for emotion expression. A renowned psychologist and professor emeritus, Paul Ekman, at the University of California, San Francisco, advocated the six basic human moods classification: happiness, sadness, disgust, fear, surprise, and anger [32] [33]. Ekman later added "contempt" as the seventh primary human emotion to his list [34] [35]. Another renowned psychologist, Robert Plutchik, invented the Wheel of Emotions, advocated eight primary emotions: anger, anticipation, joy, trust, fear, surprise, sadness, and disgust [13]. Research at Glasgow University in 2014 was amended that couple pairs of primary human emotions, such as fear and surprise, elicit similar facial muscles response, so are disgust and anger. The study broke the raw human emotions down to four fundamental emotions: happiness, sadness, fear/surprise, and disgust/anger [36] [37]. This paper adopts Paul Ekman's classification of six primary human emotions: happiness, sadness, disgust, fear, surprise, and anger for modeling the ivec embeddings while adding "neutral" as the seventh emotion feature for convenience in affective computing.

FDR's facial expression has a drawback - it fails to classify an image's emotional features with the absence of a human face. In using FDR to classify movie poster images, the poster may often contain a faceless image. Thus, we propose to indirectly classify a poster image's affective features through textual-based emotion detection and recognition (EDR) using a movie overview rather than facial-based FDR directly on the poster image.

## 4. METHODOLOGY

We derive an innovative method as our contribution to Recommender research, which based on the following sources:

- item's explicit rating information
- item's implicit affective data embeddings
- user's emotion and taste profile embedding

We want to develop a Multi-channel Emotion Aware Recommender System (McEARS). Researchers have observed that emotions playing an essential role in the human decision-making process [38] [39] [40] [41] [42] [43]. Psychologists and practitioners in social science know that the state of mind or moods of an individual affects his decision-making processes [44] [45] [46] [47]. We envision that affective embeddings can represent any product or service. In our previous work [2], we illustrated a method to derive an emotion classifier from tweets' affective tags and use the affective model to predict a movie's mood through the movie overview. We denoted the mood embeddings of the movie as mvec. We also stated that the value of the embedding of a mvec would hold the same value throughout its lifespan. We also denote uvec representing the average value of all mvec of the movies a user has watched. The value of uvec will change each time the user watches a movie. We want to expand the coverage of the mvec to other metadata of the movie, such as genres. We denote the expanded mvec as item embeddings (ivec), which holds a movie overview and mood embeddings. Similarly, uvec will expand its embedding as the average value of all ivec of the movies a user has consumed. We denote the expanded uvec embeddings as wvec.

### 4.1. Overview of the Tweets Affective Classifier Model

We developed the Tweets Affective Classifier (TAC), as illustrated in [2], which employed an asymmetric butterfly wing double-decker bidirectional LSTM - CNN Conv1D architecture to detect and recognize emotional features from tweets' text messages. We have preprocessed the seven emotion words embeddings to be used as input to train TAC through the pre-trained GloVe embeddings using the glove.twitter.27B.200d.txt dataset. We have two types of input words embeddings: trainable emotion words embeddings and frozen emotion words embeddings. By frozen the embeddings, we mean the weights in the embeddings are frozen and cannot be modified during TAC's training session. We started with the first half of the butterfly wing by feeding preprocessed TAC input emotion words embeddings to the double-decker bidirectional LSTM neural nets. We fed the frozen emotion word embeddings to the top bidirectional LSTM and fed trainable emotion word embeddings to the bottom bidirectional LSTM. Next, we concatenated the top and bottom bidirectional LSTM to form the double-decker neural net. We fed the output from the double-decker bidirectional LSTM to seven sets of CNN Conv1D neural nets with the dropout parameter set at 0.5 in each set of Conv1D as regularization to prevent the neural net from overfitting. We then concatenated all the outputs of Conv1Ds to form the overall output of the first half of the butterfly wing neural nets.

The architecture layout of the second half of the butterfly wing neural nets is different from the peer. We started by setting up seven pairs of CNN Conv1D neural nets. With each pair of Conv1D, we fed in parallel the preprocessed TAC's frozen emotion words embeddings as input to a Conv1D and the trainable emotion words embeddings to the other. We set the dropout parameter at 0.5 for all seven pairs of conv1D to prevent overfitting. We concatenated all the outputs of seven pairs Conv1D to become a single output and fed that to a single bidirectional LSTM with the dropout value set at 0.3. We then concatenated the first half of the butterfly wing output with the second half to form the overall output. Next, the output is then fed through in series to a MaxPooling1D with the dropout value set at 0.5, followed by a Flatten neural net before going through a Dense neural net and another Dense neural net sigmoid activation to classify the emotion classification in a probabilistic distribution. When predicting a movie's emotion profile using TAC, the classifier will classify a movie's mood through the movie overview. TAC output the movie emotion prediction in the form of the probabilistic distribution of seven values, indicating the value in percentage of each class of the seven emotions.

### 4.2. Overview of Comparative Platform for Recommenders

Building the comparative platform for Recommenders from scratch provides a way to study and observe the process of making recommendations under different context situations. We apply the most basic method to build the collection of Recommenders in the comparative platform. Thus, we are not aiming for best practice algorithms to build Recommender with high performance or high throughput in mind; but it is easy to modify and adapt to a different information context and highly functional and most desirable. A Recommender is known to build with a specific domain in mind. As we march down the path of researching Emotion Aware Recommenders, we want the comparative platform that we are developing for the movie-oriented Recommenders can later transfer the learning to other information domains.

We reckon that in the context of the movie domain, for example, a Genres Aware Recommender (GAR) may be adequate for making movie recommendations through movie genres, but without some adaptable in processing logic, the movie GAR may not handle well when feeding it with music genres. Of course, movie GAR will fail to make recommendations if we feed other domain data absence of genre information. However, primary human emotions are the same universally in different races and cultures. Once we obtain an emotion profile of a user obtains from a domain, the user's same emotion profile should be transferable to other domains with no required modification. The caveat is that the other domain must contain emotion detectable and recognizable or emotion-aware enable.

### 4.3. Datasets

The success of any machine learning project requires large enough domain-specific data for computation. For movie-related affecting computing, no affective labeled dataset is readily available. Thus, we need to build the required dataset by deriving it from the following sources. For movie rating datasets, we obtained these datasets from the GroupLens' MovieLens repository [48]. We scraped The Movie Database (TMDb) [48] for movie overviews and other metadata. MovieLens contains a "links" file that provides cross-reference links between MovieLens' movie id, TMDb'stmdb id, and IMDb's imdb id. We connect MovieLens and TMDb datasets through the "links" file.

Applying a brute force method, we scrape the TMDb database for movie metadata, particularly for movie overview or storyline, which contains subjective writings of movie descriptions that we can classify the text's mood. We can query the TMDb database by tmdb id, a unique movie identifier assigned to a movie. The tmdb id starts from 1 and up. However, in the sequence of tmdb id, gaps may exist between consecutive numbers. Our scraping effort yields 452,102 records after the cleansing of raw data that we scraped from TMDb.

We developed a seven text-based emotion classifier capable of classifying seven basic human emotions in tweets, as illustrated in [2]. We apply the Tweets Affective Classifier (TAC) to classify movie overviews' moods by running TAC through all the 452,102 overviews that scraped from the TMDb database to create a movie emotion label dataset.

MovieLens datasets come in different sizes. We work with the following MovieLens datasets: the ml-20m dataset, 20 million rating information; the ml-latest-small dataset, about ten thousand rating information of 610 users; ml-latest-full dataset, holds 27 million rating information; and the recently leased ml-25m dataset, with 25 million rating information. The MovieLens dataset's name coveys the number of ratings, movies, users, and tagsin the dataset. Table 1 depicts the number of ratings, users, and movies; each of the MovieLens datasets contain. Each of the depicted MovieLens datasets provides a links file to cross-reference between MovieLens and two other movie databases, TMDb and the Internet Movie Database (IMDb ) [50], by using tmdb id, imdb id, and movie id. Users can link MovieLens to TMDb and IMDb databases via the links file to access other metadata that MovieLens lacks.

**Table 1:** MovieLens datasets.

| Datasets | Ratings | Users | Titles |
|---|---|---|---|
| ml-20m | 20000263 | 138493 | 27278 |
| ml-25 | 25000085 | 162541 | 62423 |
| ml-latest-small (mlsm) | 100836 | 610 | 9742 |
| ml-latest-full {ml27m} | 27763444 | 283228 | 58098 |

The ml-latest-full dataset is the largest in the MovieLens dataset collection. However, the ml-latest-full dataset will change over time and is not proper for reporting research results. We use the ml-sm and ml-27m datasets in proof of concept and prototyping, not research reporting work. The other MovieLens 20M and 25M datasets are stable benchmark datasets which we will use for research reporting.

One drawback of MovieLens datasets is that user id does not consistently maintain across different MovieLens datasets but maintain consistently within a dataset. In other words, the same user id that appears in different MovieLens datasets may not be the same user but only is the same user when it appears within a MovieLens dataset. For example, the user id that appears in MovieLens ml-latest-small dataset is not the same user as appears in the ml-20m dataset.

International Journal on Natural Language Computing (IJNLC) Vol.9, No.4, August 2020

We propose interconnecting the disjoint users across different MovieLens datasets using an innovative Affective Aware Pseudo Association (AAPA) method. We have developed in [2] a method to classify movies' emotion profiles, mvec, through a Tweets Affective Classifier over movie overviews. We compute a user's emotion profile, uvec, by taking the average of all movies the user has watched. We compute uvec for all users in the ratings.csv datafile containing in the four MovieLens datasets. Knowing that users are not consistent across MovieLens datasets, for instance, the user with user id 400 in the ml-latest-small dataset is not the same user whose user id is 400 in ml-20m or ml-25m or ml-latest-full datasets. To associate the ml-latest-small dataset's user id 400 to other MovieLens' datasets, we compute the Cosine Similarity pairwise between ml-latest-small user id 400's uvec with other MovieLens datasets users' uvec. We pseudo associate the user in other MovieLens datasets whose uvec is most similar to the uvec of user id 400 in the ml-latest-small dataset. Table 2 depicts the pseudo associate connection (PAC) of the ml-latest-small dataset user id 400 to users in other MovieLens datasets whose uvec are most similar to the uvec of the user id 400.

A careful examination of the (PAC) of user id 400 in the ml-latest-small dataset or denoted as mlsm to the other three larger MovieLens datasets as depicted in Table 2, user id 400 in mlsm PAC to user id 66274 in ml20, user id 95459 in ml25m, and user id 89185 in ml-latest-full or denoted as ml27m. Cosine Similarity between user id 400 in mlsm and other PAC users are virtually identical. In fact, except for PAC users in ml20m, the other PAC users in ml25m and ml27m are identical. Both PAC users in ml25m and ml27m have the same history of movie watched list as user id 400 in mlsm. Thus, their uvecs are perfectly identical. Therefore, we deduce that all said three users are the same user in the respective MovieLens dataset. We can conclude that the AAPA scheme works. The method provides a means to associate disjointed users across different datasets. However, in our case, reconnecting the disjointed users for user id 400 in mlsm to other PAC users does not offer any additional values because those PAC disjointed users do not contain any new information value to user id 400. We fail to enlarge any extra data point for user id 400 in mlsm through PAC to other disjointed users in other MovieLens datasets.

Instead of selecting a low movie watched count user such as user id 400 in mlsm only has watched 43 movies, we decide to select users with a high movie watched count as test users. In the mlsm dataset, we found the following users have movie watched count in 1,300 to 2,700. Depicted in Table 2 are newly selected test users and their PAC information. Table 2 depicted five columns. The first column depicted the identity of user 1 to user 4. The role they play for the specific row information. For example, after the label row, row 1 shows user 1 who user id is 400 in the mlsm dataset, it can PAC to user id 66274 of the ml-20m dataset, or can PAC from mlsm to user id 85459 in the ml-25m dataset. Finally, it also can PAC to user id 89195 in the ml27m dataset. There are rows contain rated movies counts for each respective user in different datasets. There are rows holding movies watched list, uvecs, and values of Cosine Similarity of respective users. Table 2: Pseudo associate ml-latest-small users with high movie watched count.

**Table 2:** Pseudo associate ml-latest-small users with high movie watched count.

| Data set | mlsm | ml20m | ml25m | ml27m |
|---|---|---|---|---|
| User1PAC | 400 | 66274 | 95459 | 89195 |
| User2PAC | 414 | 125022 | 131662 | 236165 |
| User3PAC | 448 | 63555 | 134534 | 182133 |
| User4PAC | 474 | 96370 | 107581 | 54271 |
| User1Movie Count | 43 | 22 | 43 | 43 |

| User2 Movie Count | 2698 | 694 | 694 | 2698 |
|---|---|---|---|---|
| User3 Movie Count | 1864 | 290 | 1842 | 1863 |
| User4 Movie Count | 2108 | 2108 | 434 | 2108 |
| User1Watch List | [6, 47, 50, 260, …, 122886, 134130, 164179, 168252] | [47, 260, 300, 307, ...,2628, 2797, 3418, 3481] | [6, 47, 50, 260, …, 122886, 134130, 164179, 168252] | [6, 47, 50, 260, …, 122886, 134130, 164179, 168252] |
| User2Watch List | [1, 2, 3, 5, …, 180497, 180985, 184791, 187595] | [1, 2, 6, 10, …, 38886, 41716, 42004, 44191] | [1, 2, 6, 10, …, 38886, 41716, 42004, 44191] | [1, 2, 3, 5, …, 180497, 180985, 184791, 187595] |
| User3Watch List | [1, 2, 3, 5, …, 168350, 168456, 169180, 173751] | [10, 19, 145, 231, …, 117590, 118082, 120635, 128488] | [1, 3, 4, 5, …, 168350, 168456, 169180, 173751] | [1, 2, 3, 5, …, 168350, 168456, 169180, 173751] |
| User4Watch List | [1, 2, 5, 6, …, 56563, 56607, 63433, 66934] | [1, 2, 5, 6, …, 56563, 56607, 63433, 66934] | [1, 2, 6, 7, …, 41566, 45499, 45722, 49272] | [1, 2, 5, 6, …, 56563, 56607, 63433, 66934] |
| User 1 uvec | [0.16352993 0.08873525 0.12708998 0.2033184 0.11933819 0.15881287 0.13917538] | [0.16250185 0.08608596 0.12653955 0.20701054 0.11776195 0.16004661 0.14005356] | [0.16352992 0.08873525 0.12708998 0.2033184 0.1193382 0.15881287 0.13917539] | [0.16352993 0.08873526 0.12708998 0.20331841 0.11933819 0.15881286 0.13917538] |
| User 2 uvec | [0.16635188 0.09730581 0.1180924 0.1641951 0.11517799 0.17250315 0.16637367] | [0.16651482 0.09717311 0.11856494 0.16367588 0.1142451 0.17262486 0.16720128] | [0.16651482 0.09717311 0.11856494 0.16367588 0.11424511 0.17262486 0.16720128] | [0.16635188 0.09730581 0.1180924 0.1641951 0.11517799 0.17250315 0.16637366] |
| User 3 uvec | [0.17283309 0.09685813 0.11604573 0.16120733 0.11227607 0.17098578 | [0.17277665 0.097632 0.11662544 0.1612837 0.11214512 0.17031239 | [0.17312124 0.0967076 0.11600515 0.16130038 0.11236504 0.17084345 | [0.17283749 0.09688241 0.11604817 0.16112697 0.11227325 0.17099048 |

|  | 0.16979389] | 0.16922469] | 0.16965714] | 0.16984123] |
|---|---|---|---|---|
| User4uvec | [0.16885831 0.09974659 0.1187206 0.16087716 0.11261272 0.17191968 0.16726495] | [0.16885831 0.09974659 0.1187206 0.16087715 0.11261272 0.17191968 0.16726495] | [0.16932797 0.09930285 0.11946991 0.16035437 0.11281711 0.17219671 0.16653107] | [0.16885831 0.09974659 0.1187206 0.16087716 0.11261272 0.17191968 0.16726494] |
| Cossim1 | 1.0 | 0.9999153746925892 | 0.9999999999999992 | 0.9999999999999994 |
| Cossim2 | 1.0 | 0.9999929502005598 | 0.9999929501993227 | 1.0 |
| Cossim3 | 1.0 | 0.9999943241759192 | 0.9999994563170516 | 0.9999999687464807 |
| Cossim4 | 1.0 | 0.9999999999999998 | 0.9999935802799591 | 1.0 |

Although from TMDb, we have scraped 452,102 movie overviews when merging with MovieLens, we can only use one-eighth of the number of overviews that we have collected. Table 3 shows the number of movie overviews the MovieLens datasets can extract from TMDb after cleaning from raw data.

**Table 3:** Number of overviews in MovieLens extracted from TMDb.

| **Datasets** | **No. of Overviews** |
|---|---|
| ml-20m | 26603 |
| ml-25m | 25M |
| ml-latest-small | 9625 |
| ml-latest-full | 56314 |

We merged the MovieLens datasets with the emotion label datasets obtained from TAC. Form our cleansed ml-latest-small training dataset of 9625 rows extracted from the raw 9742 rows, after merging with the emotion label dataset, the applicable data point row is down to 9613. MovieLens datasets are known for preprocessed and cleaned datasets. Nevertheless, when going through the necessary data preparation steps, we still experienced a 1.32% data loss from the original dataset. Depicted below in Table 4 is the first few rows of the final cleansed training dataset.

**Table 4:** First few rows of the cleansed training dataset.

| **Index** | **tid** | **mid** | **iid** | **mood** | **neutral** | **happy** | **sad** | **hate** | **anger** | **disgust** | **surprise** |
|---|---|---|---|---|---|---|---|---|---|---|---|
| 1 | 2 | 4470 | 94675 | disgust | 0.157 | 0.086 | 0.156 | 0.075 | 0.085 | 0.266 | 0.175 |
| 2 | 5 | 18 | 113101 | disgust | 0.121 | 0.060 | 0.098 | 0.128 | 0.133 | 0.244 | 0.216 |
| 3 | 6 | 479 | 107286 | hate | 0.075 | 0.114 | 0.054 | 0.433 | 0.095 | 0.128 | 0.100 |
| 4 | 11 | 260 | 76759 | neutral | 0.299 | 0.262 | 0.079 | 0.030 | 0.017 | 0.083 | 0.230 |
| 5 | 12 | 6377 | 266543 | surprise | 0.150 | 0.080 | 0.055 | 0.083 | 0.103 | 0.153 | 0.376 |

## 5. IMPLEMENTATION

### 5.1. Recommender Platform

We develop a movie Recommender platform for our study to evaluate five Recommender algorithms in movie recommendations making. We employ the following five Recommender algorithms in the Recommender platform.

- an Item-based Collaborative Filtering (IBCF) movie Recommender to compute pairwise items Cosine Similarity, as depicted in equation 2, identifies the closeness of similar items. The rating matrix, *R*, configure with rows representing movie titles and columns representing users.
- a User-based Collaborative Filtering (UBCF) movie Recommender to compute pairwise users Cosine Similarity, as depicted in equation 2, identifies the closeness of similar users. The rating matrix, *R*, configure with rows representing users and columns representing movie titles.
- A genre-aware Content-based Recommender (GAR) using Cosine Similarity as depicted in equation 2 to compute the pairwise similarity between two movies' genres.
- an emotion aware Content-based Recommender (EAR) using Cosine Similarity as defined in equation 2 to compute the pairwise similarity between two emotion aware movies.
- an emotion and genres aware multi-modal Content-based Recommender (MAR) using Cosine Similarity as depicted in equation 2 to compute the pairwise similarity between two items with affective awareness and genres embeddings.
-

$$Inner(x, y) = \sum_i x_i y_i = <x, y> \qquad (1)$$

$$CosSim(x, y) = \frac{\sum_i x_i y_i}{\sqrt{\sum_i x_i^2} \sqrt{\sum_i y_i^2}} = \frac{<x, y>}{||x|| ||y||} \qquad (2)$$

We deployed the MovieLens ml-latest-small dataset as the training set and randomly pick a user, user id 400, as the active test user. Before evaluating wvec and ivec, we prepare each user's wvec by computing the average of all ivec of the movies the user has watched. The wvec of the active test user, user id 400, depicts in Table 5, representing the overall average of 43 movies' ivec the user id 400 has watched.

**Table 5:** Test user wvecby taking average mood values from all movies he has watched.

| mls | neutral | joy | sadness | hate | anger | disgust | surprise |
|-----|---------|---------|------------|-----------|---------|----------|----------|
| 400 | 0.163529 | 0.08873 | 0.12708998 | 0.2033184 | 0.11933 | 0.1588128 | 0.13917 |
| 414 | 0.166351 | 0.09730 | 0.1180924 | 0.1641951 | 0.11517 | 0.1725031 | 0.16637 |
| 474 | 0.168858 | 0.09974 | 0.1187206 | 0.1608771 | 0.11261 | 0.1719196 | 0.16726 |
| 448 | 0.172833 | 0.09685 | 0.11604573 | 0.1612073 | 0.11227 | 0.1709857 | 0.16979 |

## 6. EVALUATION

We deployed the ml-latest-small dataset as the training dataset. We then randomly picked user id 400 as the active test user. We created the testing dataset from concatenating the other MovieLens datasets: ml-20m, ml-25m, and ml-latest-full. We extracted all data points belonging to the user id 400 and removed all the duplicated data points from the testing dataset and those found in the training dataset and named it as test id 400 dataset. The list of movies in the test id 400 dataset represents the movies the active user id 400 has yet watched. We compare the top-20 movie list generated by the Recommender algorithms against the active user unseen movie list in the testing dataset. We also get each top-5 list from each top-20 list by computing the closest similarity between the active user's wvec and each movie's ivec on the top-20 list and sorted in the descending order. In the top-5 list, films indicate a high probability the active user may accept one of the movies from the recommendations. However, the assumption we make on the active user choosing one of the recommendations' unwatched films has a drawback. If a movie the active user likes to watch does not appear on the list, he would not choose the cinema but wait till he sees the popular film shows up on the recommendation list.

### 6.1. Top-20 Lists Generated by Recommenders

To generate movie recommendations for the active test user id 400, we chose a watched movie from the user's watched list, "Indiana Jones and the Last Crusade (1989)" as a basis. For each Recommender algorithm in the platform, we generated a top-20 movie recommendations list for the user id 400. We depicted in Table 6, a collection of five top-20 movie recommendations lists made by each Recommender. Due to the limitation of space, table 4 will only show the top-20 list by movie id. For the corresponding movie titles, please refer to Table 8(A) and (B).

**Table 6.** Top-20 recommendations list generated by five recommenders for test users.

| U400 | IBCF | UBCF | GAR | EAR | MAR |
|---|---|---|---|---|---|
| 1 | 1198 | 5952 | 761 | 7386 | 2879 |
| 2 | 2115 | 6016 | 90403 | 3283 | 112897 |
| 3 | 1196 | 4226 | 112897 | 3174 | 3283 |
| 4 | 1210 | 2329 | 32511 | 35807 | 5803 |
| 5 | 1036 | 2858 | 4367 | 6911 | 25946 |
| 6 | 1240 | 1089 | 160563 | 2243 | 2471 |
| 7 | 260 | 2762 | 115727 | 1496 | 1801 |
| 8 | 1270 | 68157 | 7925 | 7132 | 7302 |
| 9 | 2716 | 48394 | 1049 | 106920 | 122918 |
| 10 | 1200 | 110 | 3999 | 2017 | 2990 |
| 11 | 1214 | 2028 | 91485 | 144478 | 95510 |
| 12 | 1580 | 1682 | 147662 | 95441 | 5540 |
| 13 | 2571 | 115713 | 50003 | 43556 | 69278 |
| 14 | 589 | 1206 | 5244 | 96530 | 7248 |
| 15 | 1527 | 1704 | 112911 | 378 | 31923 |
| 16 | 1265 | 1 | 131714 | 2248 | 149406 |
| 17 | 1097 | 3147 | 3389 | 5088 | 134775 |
| 18 | 1136 | 1732 | 704 | 5667 | 112175 |
| 19 | 2028 | 27773 | 1606 | 2879 | 79695 |
| 20 | 1197 | 1228 | 4565 | 5803 | 5264 |

| Hit % | 40% | 0% | 0% | 0% | 0% |
|---|---|---|---|---|---|
| **U414** | **IBCF** | **UBCF** | **GAR** | **EAR** | **MAR** |
| 1 | 1291 | 1258 | 73499 | 858 | 131714 |
| 2 | 1196 | 8368 | 3030 | 53519 | 129229 |
| 3 | 260 | 2424 | 4565 | 363 | 704 |
| 4 | 1270 | 1230 | 3389 | 3109 | 112911 |
| 5 | 1210 | 1982 | 1049 | 3330 | 1606 |
| 6 | 2115 | 3176 | 2153 | 6981 | 3389 |
| 7 | 2571 | 1219 | 809 | 3211 | 2421 |
| 8 | 1240 | 48385 | 7925 | 5569 | 3704 |
| 9 | 1197 | 34542 | 131714 | 5853 | 442 |
| 10 | 1036 | 1449 | 112897 | 4224 | 4367 |
| 11 | 1136 | 6879 | 27837 | 4608 | 87430 |
| 12 | 1200 | 42418 | 1867 | 135133 | 2826 |
| 13 | 1214 | 520 | 1606 | 109673 | 1049 |
| 14 | 2716 | 1183 | 704 | 2009 | 122922 |
| 15 | 4993 | 2788 | 56775 | 110655 | 64695 |
| 16 | 1265 | 1693 | 91485 | 29 | 2683 |
| 17 | 2028 | 40732 | 115727 | 4433 | 3702 |
| 18 | 7153 | 2324 | 32511 | 2256 | 131739 |
| 19 | 3578 | 1333 | 6664 | 3028 | 76743 |
| 20 | 858 | 2728 | 761 | 2450 | 57326 |
| Hit % | 90% | 65% | 20% | 35% | 30% |
| **U448** | **IBCF** | **UBCF** | **GAR** | **EAR** | **MAR** |
| 1 | 2922 | 293 | 76091 | 193579 | 2023 |
| 2 | 5049 | 142488 | 48883 | 3372 | 68554 |
| 3 | 2338 | 2028 | 118354 | 77364 | 8773 |
| 4 | 3441 | 364 | 68554 | 421 | 6214 |
| 5 | 40959 | 166528 | 52604 | 4721 | 507 |
| 6 | 3802 | 318 | 44759 | 90403 | 71033 |
| 7 | 6812 | 93840 | 1805 | 5282 | 90528 |
| 8 | 50792 | 34405 | 5506 | 27478 | 118354 |
| 9 | 5636 | 176371 | 628 | 166534 | 51357 |
| 10 | 1616 | 168252 | 1422 | 162478 | 88593 |
| 11 | 6314 | 88129 | 3262 | 1005 | 259 |
| 12 | 2051 | 2081 | 100383 | 1097 | 4383 |
| 13 | 1023 | 63082 | 7305 | 8952 | 165347 |
| 14 | 78467 | 3275 | 103449 | 2060 | 1834 |
| 15 | 2126 | 110 | 183011 | 8535 | 127136 |
| 16 | 3791 | 7361 | 48142 | 4626 | 61011 |
| 17 | 1614 | 44195 | 1406 | 95182 | 61352 |
| 18 | 3249 | 2329 | 4056 | 26913 | 2688 |
| 19 | 5462 | 48774 | 26002 | 48319 | 6086 |
| 20 | 2090 | 5995 | 6214 | 69860 | 68347 |
| Hit % | 90% | 65% | 20% | 35% | 30% |
| **U474** | **IBCF** | **UBCF** | **GAR** | **EAR** | **MAR** |

| 1 | 2478 | 32587 | 4215 | 3143 | 60408 |
|---|------|-------|------|------|-------|
| 2 | 2072 | 4011 | 141 | 86190 | 93270 |
| 3 | 2796 | 1527 | 163653 | 1199 | 467 |
| 4 | 3388 | 553 | 5572 | 2052 | 57532 |
| 5 | 2374 | 3189 | 1516 | 6225 | 4732 |
| 6 | 3039 | 4816 | 3048 | 159779 | 67997 |
| 7 | 414 | 2985 | 103819 | 134847 | 134847 |
| 8 | 2169 | 10 | 6425 | 40946 | 785 |
| 9 | 2015 | 3275 | 1753 | 7266 | 93980 |
| 10 | 1665 | 3 | 3861 | 72142 | 3544 |
| 11 | 2416 | 3703 | 5500 | 6062 | 168846 |
| 12 | 2615 | 5388 | 3911 | 58332 | 3911 |
| 13 | 2458 | 4865 | 148632 | 5784 | 4524 |
| 14 | 2470 | 2948 | 429 | 96079 | 830 |
| 15 | 3524 | 60069 | 4452 | 491 | 1390 |
| 16 | 2408 | 35957 | 27478 | 141408 | 1855 |
| 17 | 4564 | 5679 | 53956 | 68597 | 79428 |
| 18 | 4002 | 58559 | 8528 | 100397 | 3511 |
| 19 | 1983 | 1127 | 59429 | 7730 | 112788 |
| 20 | 4215 | 5010 | 3031 | 5590 | 120919 |
| **Hit %** | 95% | 85% | 0 | 35% | 5% |

### 6.2. Top-5 Lists Extracted from Recommenders' Top-20 Lists

Using the wvec of the user id 400, we compute the pairwise similarity between the user id 400 and each recommended movie's ivec on the top-20 list. We sorted the pairwise distance metrics top-20 list calculated using wvec in the descending order to obtain the top-5 list for each of the Recommenders. Table 7 shows the computed top-5 list for each Recommender.

**Table 7.** Top-5 list computed via wvec of test usersand ivec of top-20 corresponding Recommenders.

| U400 | IBCF | UBCF | GAR | EAR | MAR |
|------|------|------|-----|-----|-----|
| 1 | 2716 | 1732 | 90403 | 3174 | 122918 |
| 2 | 1527 | 2329 | 1606 | 43556 | 134775 |
| 3 | 2115 | 2858 | 5244 | 95441 | 112175 |
| 4 | 1240 | 6016 | 4367 | 5667 | 25946 |
| 5 | 1036 | 1206 | 147662 | 144478 | 79695 |
| **Hit %** | 80% | 0% | 0% | 0% | 0% |
| **U414** | **IBCF** | **UBCF** | **GAR** | **EAR** | **MAR** |
| 1 | 3578 | 2728 | 1606 | 2009 | 1606 |
| 2 | 2716 | 6879 | 56775 | 110655 | 4367 |
| 3 | 2115 | 2788 | 3030 | 363 | 2421 |
| 4 | 1240 | 2324 | 7925 | 3109 | 1049 |
| 5 | 1036 | 520 | 1049 | 4224 | 704 |
| **Hit %** | 100% | 80% | 40% | 20% | 20% |
| **U448** | **IBCF** | **UBCF** | **GAR** | **EAR** | **MAR** |
| 1 | 5636 | **176371** | 5506 | 95182 | 51357 |

| 2 | 5049 | 5995 | 1422 | 26913 | 259 |
| 3 | 3441 | 2329 | 1805 | 1005 | 68554 |
| 4 | 3791 | 34405 | 100383 | 90403 | 68347 |
| 5 | 1614 | 48774 | 1406 | 4721 | 507 |
| **Hit %** | 100% | 80% | 40% | 20% | 20% |
| **U474** | **IBCF** | **UBCF** | **GAR** | **EAR** | **MAR** |
| 1 | 2458 | 1527 | 4452 | 40946 | 57532 |
| 2 | 2478 | 10 | 5572 | 141408 | 93270 |
| 3 | 3039 | 2985 | 163653 | 5590 | 3911 |
| 4 | 1983 | 35957 | 3861 | 7266 | 4524 |
| 5 | 2470 | 1127 | 141 | 6062 | 79428 |
| **Hit %** | 100% | 100% | 0% | 60% | **0%** |

### 6.3. Limitations

The Affective Aware Pseudo Association (AAPA) method works, and indeed it works so well that by close examination Table 2, the content of the pair of disjoint users that AAPA connects across different datasets often resemblance the characteristic of the same user. It appears that MovieLens duplicated some data points across different datasets. However, some data points are almost an exact clone between different datasets, whereas some are partial copies. In either case, no additional information value we can extract from the AAPA datasets to our intent dataset enlargement goal. Listed below are limitations that we observed in the study.

- Due to the lack of emotion labeled movie datasets, we cannot avowthe accuracy of the moods labeled movie dataset generated from the Tweets Affection Classifier. However, from our observation, TAC did a fair job of classifying films' emotional attributes from overviews.
- We adopted seven categories of emotion and not more because the more classes we want to add to the collection, the harder we can find adequate labeled data. From our observation, seven classes seem to be the limit.
- The top-N list that generates by each Recommender shares little in common. It presents a problem in finding an adequate evaluation metrics for benchmarking. The five top-20 recommendations lists made by each respective Recommender are so different that we find ourselves comparing these lists as if comparing apples and oranges. They are all fruits but with very different tastes. For the time being, we rely on our intuition to judge how good these top-20 and top-5 lists are.
- One objective metrics we apply to evaluate the performance of Top-N is to withhold a percentage of data points for testing and validation from training. For example, we sorted all test users' data points against the timestamp attribute in the rating data file to stimulate the testing user's movie-watching order. We split the data points into training (20\%) as the user watched the movies list, testing, and validation (80\%) for movies to be watched. We count and report the number of movies matching between the to-be-watched list and the Top-N list as hit percentages.

## 7. FUTURE WORK

We started the track to study the impact of affective features have on Recommender Systems by examining how emotional attributes can interplay at the stage of the Recommender, making top-N recommendations in [2]. Next, we elaborate on the work in [1] by demonstrating how to make Recommender emotionally aware. We focused on extracting affective features from textual movie metadata. This paper introduces an innovative affective aware pseudo association method (PAM) to reconnect disjoint users across different datasets. We plan on an in-depth study of emotion-aware recommendation-making in a Multi-channel Emotion Aware Recommender by extracting emotion features from images such as movie posters as a component in building the Recommender. We want to examine the idea of using users' emotion profiles to enhance Group Recommenders in user grouping, group formation, group dynamics, and group decision making.

Table 8 (A). TopN recommendation list generated by five recommenders for userid 400.

| No. | Mid | Title |
| --- | --- | --- |
| 0 | 1 | Toy Story (1995) |
| 1 | 110 | Braveheart (1995) |
| 2 | 260 | Star Wars: Episode IV - A New Hope (1977) |
| 3 | 378 | Speechless (1994) |
| 4 | 589 | Terminator 2: Judgment Day (1991) |
| 5 | 704 | Quest, The (1996) |
| 6 | 1036 | Die Hard (1988) |
| 7 | 1049 | Ghost and the Darkness, The (1996) |
| 8 | 1089 | Reservoir Dogs (1992) |
| 9 | 1097 | E.T. the Extra-Terrestrial (1982) |
| 10 | 1197 | Princess Bride, The (1987) |
| 11 | 1206 | Clockwork Orange, A (1971) |
| 12 | 1210 | Star Wars: Episode VI - Return of the Jedi (1983) |
| 13 | 1214 | Alien (1979) |
| 14 | 1240 | Terminator, The (1984) |
| 15 | 1265 | Groundhog Day (1993) |
| 16 | 1270 | Back to the Future (1985) |
| 17 | 1496 | Anna Karenina (1997) |
| 18 | 1527 | Fifth Element, The (1997) |
| 19 | 1580 | Men in Black (a.k.a. MIB) (1997) |
| 20 | 1682 | Truman Show, The (1998) |
| 21 | 1704 | Good Will Hunting (1997) |
| 22 | 1732 | Big Lebowski, The (1998) |
| 23 | 1801 | Man in the Iron Mask, The (1998) |
| 24 | 2017 | Babes in Toyland (1961) |
| 25 | 2243 | Broadcast News (1987) |
| 26 | 2248 | Say Anything... (1989) |
| 27 | 2471 | Crocodile Dundee II (1988) |
| 28 | 2571 | Matrix, The (1999) |
| 29 | 2716 | Ghostbusters (a.k.a. Ghost Busters) (1984) |
| 30 | 2762 | Sixth Sense, The (1999) |

Table 8 (B). Top-N recommendation list generated by five recommenders for user id 400.

| No. | Mid | Title |
| --- | --- | --- |
| 31 | 2858 | American Beauty (1999) |
| 32 | 2990 | Licence to Kill (1989) |
| 33 | 3147 | Green Mile, The (1999) |
| 34 | 3389 | Let's Get Harry (1986) |
| 35 | 3999 | Vertical Limit (2000) |
| 36 | 4367 | Lara Croft: Tomb Raider (2001) |

| 37 | 5088 | Going Places (Valseuses, Les) (1974) |
| --- | --- | --- |
| 38 | 5244 | Shogun Assassin (1980) |
| 39 | 5540 | Clash of the Titans (1981) |
| 40 | 5667 | Tuck Everlasting (2002) |
| 41 | 6911 | Jolson Story, The (1946) |
| 42 | 7132 | Night at the Opera, A (1935) |
| 43 | 7248 | Suriyothai (a.k.a. Legend of Suriyothai, The) (2001) |
| 44 | 7302 | Thief of Bagdad, The (1924) |
| 45 | 7925 | Hidden Fortress, The (Kakushi-toride no san-akunin) (1958) |
| 46 | 25946 | Three Musketeers, The (1948) |
| 47 | 31923 | Three Musketeers, The (1973) |
| 48 | 43556 | Annapolis (2006) |
| 49 | 48394 | Pan's Labyrinth (Laberinto del fauno, El) (2006) |
| 50 | 50003 | DOA: Dead or Alive (2006) |
| 51 | 68157 | Inglourious Basterds (2009) |
| 52 | 69278 | Land of the Lost (2009) |
| 53 | 95441 | Ted (2012) |
| 54 | 96530 | Conception (2011) |
| 55 | 106920 | Her (2013) |
| 56 | 112175 | How to Train Your Dragon 2 (2014) |
| 57 | 112911 | Hercules (2014) |
| 58 | 115713 | Ex Machina (2015) |
| 59 | 115727 | Crippled Avengers (Can que) (Return of the 5 Deadly Venoms) (1981) |
| 60 | 122918 | Guardians of the Galaxy 2 (2017) |
| 61 | 131714 | Last Knights (2015) |
| 62 | 134775 | Dragon Blade (2015) |
| 63 | 147662 | Return of the One-Armed Swordsman (1969) |
| 64 | 149406 | Kung Fu Panda 3 (2016) |
| 65 | 160563 | The Legend of Tarzan (2016) |

## 8. CONCLUSION

We leveragedour prior affective computingwork in[2] to generate a movie emotion labeled dataset by the Tweets Affective Classifier (TAC). Weextended the work in [1] with a method to pseudo associatedisjoint users across different datasets,hoping to add more users' data points to build a more massivedataset for training, testing, and validation. We demonstrated using the pseudo association method (PAM)to build a massivedataset for training, testing, and validation in Recommender development. We make use of the timestamp attribute found in MovieLens datasets and sorted a test user movie watched list according to his movies watched order. By observing the hit rate on the top-N recommendations list against the movies to be watched list, we computed the coverage accuracy.

We developed a Recommender platform using the following Recommender algorithms: Item-based Collaborative Filtering (IBCF), User-based Collaborative Filtering (UBCF), Content-based movie Genres Aware Recommender (GAR), Content-based Emotion Aware Recommender (EAR), and Content-based Multi-channel Emotion Aware Recommender (MAR). With each Recommender algorithm, we generate a top-20 recommendation list. We randomly selected user id 400 as an active user for testing. We compute the emotion profile, wvec, for the test user. Using the test user's wvec and the list of ivec from each of the top-20 list, we computed the top-5 for each top-20 list generated by the Recommenders. The top-N list made by each Recommender is unique, with few overlaps among the lists. We have a total of 100 movies in the combined top-20 lists. We found 35 duplicated films among the top-20 recommendation lists. The top-N list made by each Recommender met its design focus. For example, GAR correctly recommended movies that meet the active test user's genre taste. EAR, on the other hand, shows intrigue results. We believe that with further investigation, we could enhance EAR to make serendipity recommendations.

## AUTHORS

**John K. Leung** is a Ph.D. candidate in Computational and Data Sciences Department, Computational Sciences and Informatics at George Mason University in Fairfax, Virginia. He has over twenty years of working experience in information technology research and development capacity. Formerly, he worked in the T. J. Watson Research Center at IBM Corp. in Hawthorne, New York. John has spent more than a decade working in Greater China, leading technology incubation, transfer, and new business development.

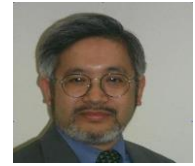

**Igor Griva** is an Associate Professor in the Department of Mathematical Sciences at George Mason University. His research focuses on the theory and methods of nonlinear optimization and their application to problems in science and engineering.

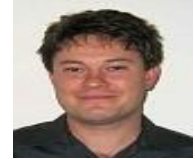

**William G. Kennedy,** PhD, Captain, USN (Ret.) is an Associate Professor in the Department of Computationaland Data Sciences and is a Co-Director of the Center for Social Complexity at George Mason University in Fairfax, Virginia. He has over 10-years' experience in leading research projects in computational social science with characterizing the reaction of the population of a mega-city to a nuclear WMD event being his most recent project. His teaching, research, and publication activities are in modeling cognition and behavior from individuals to societies.

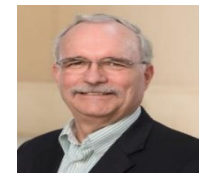